\documentclass[aps,prl,twocolumn,amsmath,amssymb,superscriptaddress,floatfix]{revtex4-1}

\usepackage{graphicx}
\usepackage{bm}
\usepackage[usenames]{color}

\bibstyle{apsrev.bib}
\newcommand{\be}{\begin{equation}}
\newcommand{\ee}{\end{equation}}
\newcommand{\beqn}{\begin{eqnarray}}
\newcommand{\eeqn}{\end{eqnarray}}

\begin{document}

\title{Optimality of spatially inhomogeneous search strategies}

\author{Karsten Schwarz}
\author{Yannick Schr\"oder}
\affiliation{Theoretical Physics, Saarland University, 
66123 Saarbr\"ucken, Germany}
\author{Bin Qu}
\author{Markus Hoth}
\affiliation{Biophysics, CIPMM,
Saarland University, 66421 Homburg, Germany}
\author{Heiko Rieger}
\affiliation{Theoretical Physics, Saarland University, 
66123 Saarbr\"ucken, Germany}
\email{h.rieger@physik.uni-saarland.de}

\date{\today}

\begin{abstract}
We consider random search processes alternating stochastically between
diffusion and ballistic motion, in which
the distribution function of ballistic motion directions varies from point to
point in space. The specific space dependence of the directional distribution
together with the switching rates between the two modes of motion establishes
a spatially inhomogeneous search strategy. 
We show that the mean first passage times 
for several standard search problems -- narrow escape, reaction partner
finding, reaction-escape -- can be minimized with a directional distribution 
that is reminiscent of the spatial organization of the cytoskeleton filaments 
of cells with a centrosome: radial ballistic transport from center to periphery and back, and ballistic transport in random directions within a concentric shell of thickness $\Delta_{\rm opt}$ along the domain boundary.
The results suggest that living cells realize efficient search strategies for various intracellular transport problems {\it economically} through a spatial cytoskeleton organization that involves radial microtubules in the central 
region and only a narrow actin cortex rather than a cell body filled with randomly oriented actin filaments.
\end{abstract}

\maketitle

First passage time problems \cite{Redner} are ubiquitous in nature,
and have been intensively studied in the last decade with applications to,
among others, chemical reaction kinetics in complex geometries \cite{NatureChem-geometry-Benichou}, gene expression and the location of specific target sites on DNA and polymer chains \cite{PRL-polymerchain-Metzler,PRL-chromatin-Benichou,PRL-DNAtraps-Benichou}, intra-cellular reaction kinetics \cite{Loverdo-etal}, animal foraging \cite{PNAS-evasiveprey-Oshanin,JSTAT-Nsearcher-Schehr,PRL-speedselective}, and in the context of 
sub-diffusion \cite{PNAS-subdiffusion-Condamin,PRL-subdiffusion-Condamin}
and self-similar media \citep{PRL-formula-Benichou,Nature-fractal-Condamin}.
Optimal search strategies aim at minimizing the mean first passage 
time (MFPT) that a searcher executing a random motion needs to find a target.
Optimization is achieved by tuning the parameters of the random motion,
like the switching rates between ballistic and diffusive motion 
in the search for a hidden target, also denoted as intermittent search,
\cite{PRL-hiddentarget-Benichou,Intermittent-search-Loverdo,Benichou-Review},
the persistence length in persistent random walks \cite{PRL-optimizingpersistent-Tejedor},
the resetting rate in random motion with stochastic resetting \cite{PRL-resetting-Majumdar,PRL-resetting-Schehr}, or the desorption rate in surface mediated diffusion \cite{PRL-surfacemediated-Diff,PRL-followingboundaries-Calandre}. Search strategies that are optimal with respect to first passage 
times are robust and also optimal with respect to cover time 
\cite{cover-time}.

Search strategies considered so far were spatially homogeneous,
meaning that parameters like transition rates or the distribution 
of directions for ballistic motion in intermittent search or persistent walks 
were identical in all points of the search domain. We 
study, for the first time, the efficiency of spatially inhomogeneous 
intermittent search strategies, for three paradigmatic search problems
in spherical domains: i) the narrow escape problem
\cite{Narrow-Escape-Cell,Narrow-Escape-Review}, in which a searcher 
has to find a small region on the boundary, ii) 
reaction kinetics enhancement \cite{Loverdo-etal},
in which a searcher has to find an immobile target,
and iii) the reaction-escape problem, in which a searcher has 
first to find a reaction partner before it can escape.

We focus on a spatial inhomogeneity of the search 
strategy that is inspired
by the spatial organization of the cytoskeleton of cells with
a centrosome \cite{Alberts}. In living cells reaction partners equipped with 
molecular motors perform a random search process with stochastic
alternations between ballistic motion along cytoskeleton 
filaments and diffusion, and therefore a paradigmatic realization 
of an intermittent search process 
\cite{Loverdo-etal,Benichou-Review}.
The spatially inhomogeneous and non-isotropic distribution of filaments
and filament directions including the attachment and detachment rates 
of the molecular motors define what we denote as a spatially
inhomogeneous intermittent search strategy.

A search strategy that idealizes the cytoskeleton structure in a 
spherical cell of radius $R$ consists of 
microtubule filaments emanating radially from the microtubule organizing center 
(MTOC) in the cell center and randomly oriented actin filaments in a cortex of width $\Delta$ underneath the plasma membrane. 
Mathematically such a filament distribution is defined by the probability density $\rho_\Omega({\bf r})$ to choose direction $\Omega$ conditional 
on the switch from the diffusive to a ballistic mode at position 
${\bf r}$ and can, for simplicity, be parameterized as follows:
\begin{equation}
\rho_\Omega({\bf r})=
\left\{
\begin{array}{ccc}
p\delta_{\Omega,\Omega'({\bf r})} + q\delta_{\Omega,\Omega'(-{\bf r})}&,&
0<|{\bf r}|<R-\Delta\\
1/4\pi&,& R-\Delta<|{\bf r}|<R
\end{array}
\right. .
\label{inhom-density}
\end{equation}
$\Omega'({\bf r})$ is the direction defined by the position vector
${\bf r}$, $p$ and $q=1-p$ are the probabilities to move 
radially outwards and inwards, respectively
(corresponding to the microtubule plus and minus direction, respectively, 
where the MTOC is assumed to be located in the origin, ${\bf r}=0$).
We assumed an isotropic distribution of directions $\rho_\Omega=1/4\pi$
in the periphery here, but our framework allows to put any
direction distribution - as for instance a non-isotropic filament 
distribution in the actin-cortex, where filaments run prevalently
along the cell membrane. For the scenarios that we study here results do
not change much, details will be published elsewhere.
Note that for $\Delta=R$
one obtains the homogeneous directional distribution $\rho_\Omega({\bf r})=1/4\pi$ considered in \cite{Loverdo-etal}.

We model the random motion of the searcher by an intermittent search
process \cite{PRL-hiddentarget-Benichou,PRL-polymerchain-Metzler,Loverdo-etal,Benichou-Review},
in which a particle performs random motion in two alternating modes:
Brownian motion with diffusivity $D$, and ballistic
motion with velocity ${\bf v}$. Transitions between the modes
occur stochastically with rates $k$ and $k'$
, which we denote in the following, in reminiscence of molecular motor
assisted transport, as attachment and detachment rate, respectively. 
The underlying stochastic process is described
by a Fokker-Planck equation for the time evolution of the 
probability distribution $P_0({\bf r},t)$ for the diffusing
particle and $P_{\Omega}({\bf r},t)$ for the particle that moves
ballistically with velocity $|{\bf v}_\Omega|$  in the direction $\Omega$:
\begin{eqnarray}
\frac{\partial P_0({\bf r},t)}{\partial t}\hspace*{-0.05cm} &=&\hspace*{-0.05cm}D\,\Delta P_0({\bf r},t)
-k P_0({\bf r},t) +k'\hspace*{-0.1cm}\int \hspace*{-0.1cm} d\Omega P_{\Omega}({\bf r},t)\label{FP-eq}\\
\frac{\partial P_{\Omega}({\bf r},t)}{\partial t}\hspace*{-0.05cm}&=&
\hspace*{-0.05cm}-\nabla\hspace*{-0.2mm} \big({\bf v}_\Omega P_{\Omega}({\bf r},t)\hspace*{-0.5mm}\big)
\hspace*{-0.03cm}+\hspace*{-0.03cm} k\rho_\Omega({\bf r})P_0({\bf r},t) \hspace*{-0.03cm}- \hspace*{-0.03cm}k'P_{\Omega}({\bf r},t)
\nonumber
\end{eqnarray}
The switching rates could also be space dependent functions
$k({\bf r})$ and $k'({\bf r})$, 
thus establishing a further extension spatially inhomogeneous
search strategies. Here we focus on constant rates.

We assume $v=|{\bf v}_\Omega|$ to be 
constant throughout this study. At time $t=0$ 
the particle starts diffusively at position ${\bf r_0}$ \cite{remark3}: 
$P_0({\bf r},t=0)=\delta(\bf r -\bf r_0)$, $P_{\Omega}({\bf r},t)=0$. Apart from the stochastic detachment 
with rate $k'$ a ballistically moving particle switches {\it automatically 
to the diffusive mode} at the MTOC ($|{\bf r}|=0$) since microtubules end there,
at the inner border of the actin cortex ($|{\bf r}|=R-\Delta$) since actin 
filaments end there, and at the cell membrane ($|{\bf r}|=R$) since all 
filaments end there.
Diffusing particles are reflected at the cell membrane ($|{\bf r}|=R$),
apart from small regions in the cell membrane in exit problems, 
where absorbing boundary conditions are applied.
In the following we use rescaled dimensionless 
spatial and temporal coordinates
$\tilde{{\bf r}}={\bf r}/R$ and $\tilde{t}=vt/R$,
and the rescaled parameters $\tilde{D}=D/vR$, $\tilde{k}=Rk/v$ and 
$\tilde{k}'=Rk'/v$. 

The efficiency of a search strategy, or a specific directional distribution,
$\rho_\Omega({\bf r})$ is measured in terms of a MFPT
with respect to the events defined by the different
search problems that we consider now.

An analytical determination of MFPTs in 2d and 3d bounded 
domains is feasible for homogeneous problems \cite{PRL-complex-Noh,PRL-boundedlattice-Condamin},
but appears currently intractable for spatially inhomogeneous 
problem like the one defined by (\ref{inhom-density}).
Here we resort to high precision numerical methods and use an 
efficient Monte Carlo algorithm developed by us in \cite{Comp-Schwarz} 
to generate stochastic processes underlying (\ref{FP-eq}),
for details see \cite{Schwarz-etal-long}.
We use of the order of $10^6$ realizations of the process
for each parameter value such that the relative statistical error 
is below 0.1\%.

{\bf Narrow escape problem:} First we consider the search 
for a specific {\it small} area on the search domain boundary, 
the so-called narrow escape problem 
\cite{Narrow-Escape-Cell,Narrow-Escape-Review}.
A concrete example in which the directional distribution 
(\ref{inhom-density}) is relevant is the directed secretion 
by immune cells which requires the formation of an immunological synapse 
\cite{Dustin-1999,Dustin-Review} and the transport of vesicles containing 
secretion material towards the synapse involving the cytoskeleton 
\cite{Griffiths1,Griffiths2}.

\begin{figure}[t]
\includegraphics[width=\columnwidth]{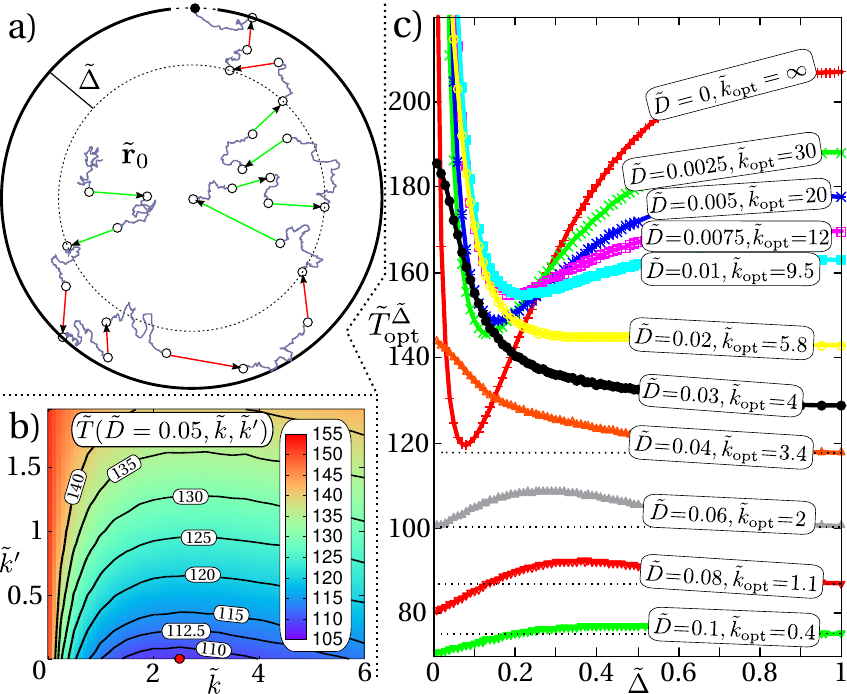}
\caption{
{\bf{a)}} Sketch of the escape process with a narrow escape region 
represented by the dotted segment on the surface of
the spherical search volume (black circle). 
Grey wiggly lines denote diffusive motion of the searcher starting at ${\bf{\tilde{r}_0}}$, 
green and red lines denote ballistic motion according to the directional
distribution (\ref{inhom-density}): in the central part 
($|{\bf \tilde{r}}|<1-\tilde{\Delta}$) only in radial direction 
(green lines),
in the periphery ($1-\tilde{\Delta}<|{\bf \tilde{r}}|<1$) in
all directions (red lines).
{\bf{b)}} 
MFPT $\tilde T$ for the
homogeneous directional distribution ($\tilde{\Delta}=1$)
in dependence of the rates $\tilde{k}$ and $\tilde{k}'$, 
exemplarily for $\tilde{D}=0.05$. The red dot indicates the optimal values (yielding the minimal 
MFPT) $\tilde k_\text{opt}(\tilde D=0.05)$ and ${\tilde{k}'_\text{opt}(\tilde D=0.05)}$.
{\bf{c)}} MFPT for the inhomogeneous directional distribution eq. (\ref{inhom-density})
as a function of $\tilde{\Delta}$ for different diffusivities  $\tilde{D}$
using the optimal rates for $\tilde{\Delta}=1$ and
$p=1$ (only outward radial transport in the central part 
($|{\bf \tilde{r}}|<1-\tilde{\Delta}$)).  
} 
\label{Fig2}
\end{figure}

We compute the MFPT for various parameters
$k$, $k'$, $p$, and $\Delta$ in (\ref{inhom-density}) and (\ref{FP-eq})  
for small escape regions of polar angle 
$\vartheta_{abso}=\text{arcsin}(1/7)\approx 0.1433$
($0.51\%$ of the spherical surface)
as sketched in Fig.\ref{Fig2}a. In order to demonstrate 
the efficiency of a spatially inhomogeneous directional distribution
(corresponding to $0<\Delta<R$ in eq. (\ref{inhom-density})) 
we first determine the optimal 
rates $\tilde k_\text{opt}(\tilde D)$ and 
${{\tilde{k}}'}_\text{opt}(\tilde D)$
for a homogeneous directional distribution ($\tilde\Delta=1$). 
Fig.\ref{Fig2}b shows the MFPT as a function of 
$\tilde{k}$ and $\tilde{k}'$ for $\tilde D=0.05$.
The optimal detachment rate $\tilde k'_\text{opt}$ is zero, 
which holds for all $\tilde D$.
$\tilde k'_\text{opt}=0$ means that uninterrupted ballistic 
transport to the cell membrane (and subsequent 
\textit{switch to the diffusive mode}) 
is optimal for the narrow escape problem,
which is plausible, since the target area is on the membrane.
The optimal attachment rate $\tilde k_\text{opt}$
decreases with increasing diffusivity  $\tilde D$
(Fig.S1a in \cite{SM}). For small diffusivities  
$\tilde D < 0.1$ the homogeneous intermittent search is 
always more efficient than the pure diffusive search.
The MFPT of the latter diverges for $\tilde{D}\to0$ 
as $\tilde T_\text{diff}\approx 7.62/{\tilde D}$,
for $\vartheta_{abso}$ as above
\cite{Cheviakov2012},
whereas the MFPT for homogeneous intermittent search 
$\tilde T_\text{opt}$ stays finite (Fig.S1a in \cite{SM}).

Next we take the optimal values $\tilde k_\text{opt}(\tilde D)$ and 
${{\tilde{k}}'}_\text{opt}(\tilde D)$ for $\tilde\Delta=1$, and 
vary the width of the cortex $\tilde\Delta$, but fix the value of $p$
in (\ref{inhom-density}) to one (only outward radial transport).
The result is shown in Fig.\ref{Fig2}c:
For small diffusivities $\tilde D<0.02$ the MFPT is minimized
for an inhomogeneous strategy $\tilde{\Delta}<0.2$.
For larger values ($0.02<\tilde D<0.06$) the 
optimal strategy for the rates $\tilde k_\text{opt}(\tilde D)$
is a homogeneous one ($\tilde{\Delta}=1$). But if we vary 
the rate $\tilde k$ superior inhomogeneous search strategies exist
also in the range $0.02<\tilde D<0.06$: Minimizing the MFPT as a function 
of $\tilde k$ and $\tilde \Delta$ simultaneously the optimal
strategy has a small, non-vanishing value for $\tilde\Delta$
(Fig.S1b in \cite{SM}) for $\tilde D=0.04$. 
For $0.06<\tilde D<0.1$ a search strategy with a fully polarized network 
($\tilde{\Delta}=0$) is optimal, and even more efficient general 
distributions $\rho_\Omega({\bf r})$ for this case exist.
Finally, also for fixed attachment and detachment rates, $\tilde k$ and $\tilde k'$,
the MFPTs are minimized for small values of $\tilde\Delta$
(Fig.S1c in \cite{SM}).

{\bf Enhanced reaction kinetic:} Next we consider the enhancement of the reaction 
kinetics between two reaction partners by motor assisted ballistic transport.
It has already been demonstrated that spatially homogeneous 
and isotropic intermittent search strategies can decrease 
MFPTs substantially \cite{Loverdo-etal,Benichou-Review}.
Such intermittent search strategies are only
realized in those parts of a biological cell, where cytoskeleton
filaments are homogeneously and isotropically distributed, which is
certainly not true for the whole cell body for which a 
spatial organization as defined by (\ref{inhom-density})
is more realistic. In the following we study the efficiency 
of this strategy for the search for an immobile reaction partner that is 
preferentially located in a specific sub-volume of the search domain
\cite{remark}.
\begin{figure}[t]
\includegraphics[width=\columnwidth]{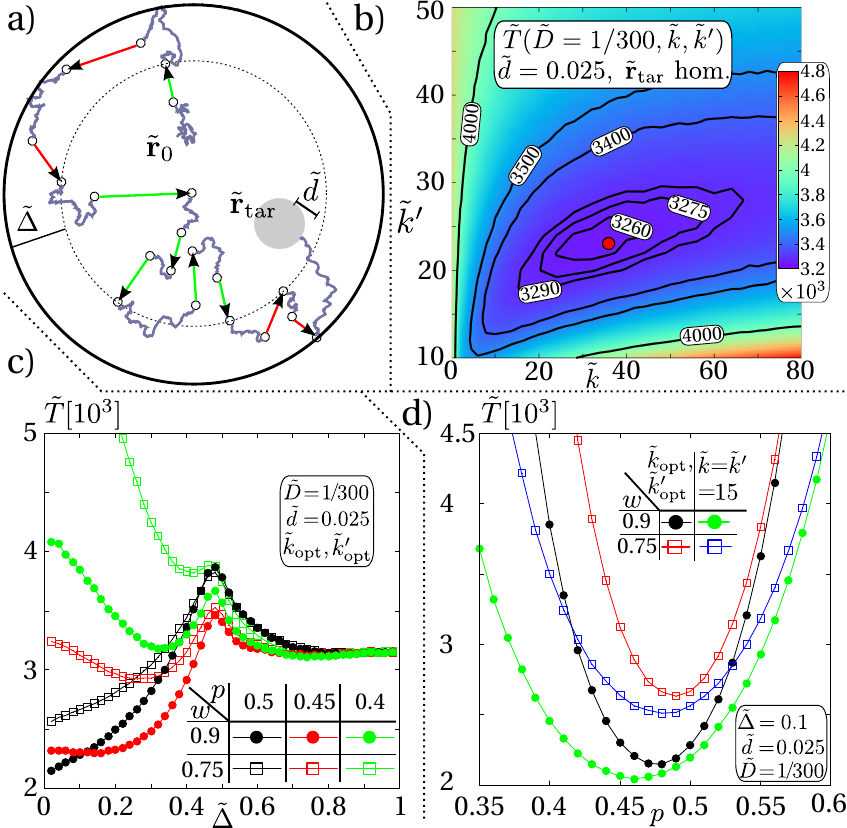}
\caption{ 
{\bf{a)}} Sketch of the intermittent search for an immobile target of
diameter $\tilde{d}$ indicated by the grey circle. Trajectories are represented 
as in Fig.\ref{Fig2}a. 
{\bf{b)}} 
$\tilde T$ as a function of $\tilde k$ and $\tilde k'$ for a spatially homogeneous directional distribution ($\tilde{\Delta}= 1$) 
and a homogeneously distributed target position $\tilde{\bf r}_\text{tar}$ for $\tilde D= 1/300, \tilde d= 0.025$.
{\bf{c)}} $\tilde T$ for the inhomogeneous directional distribution
with the optimal rates $\tilde k_\text{opt}(\tilde D, \tilde d)$
from the homogeneous case $\tilde{\Delta}= 1$ as function of 
$\tilde{\Delta}$ for different values of the forward radial transport $p$
and different target positions close to the center $w$.
{\bf{d)}} $\tilde T$ as in c) but now with fixed width $\tilde{\Delta}= 0.1$ 
as function of the forward probability $p$ for different 
fixed rates $\tilde k$ and $\tilde k'$ different values of $w$.
}
\label{Fig3}
\end{figure}
Fig.\ref{Fig3}a shows a sketch of the process for the case of an immobile target
at position $\tilde {\bf r}_\text{tar}$. When the searcher is in the diffusive mode and 
its position $\tilde {\bf r}$ comes closer to the target than 
$|\tilde {\bf r}-\tilde {\bf r}_\text{tar}|\leq \tilde d$ the search is successfully finished. The target position 
$\tilde {\bf r}_\text{tar}$ with $0\leq\tilde {\bf r}_\text{tar}\leq
1-\tilde d$ will be either homogeneously distributed or it is predominantly located close 
to the center ${\bf r}_\text{tar} \leq1/2$ with probability $w$. 

Fig.\ref{Fig3}b shows the MFPT for the homogeneous 
direction distribution ($\tilde{\Delta}=1$) 
and the parameters $\tilde D=1/300$ and $\tilde d=0.025$ 
as a function of the rates $\tilde{k}$ and $\tilde{k'}$.  
In contrast to the narrow escape problem 
the optimum is not at $\tilde{k'}= 0$ any more, 
which is plausible since the target is not located on the boundary.
The concentration of the target towards the center ($w=0.9$) does not 
change the MFPT significantly even for a rather large value of 
$\tilde d$ (see Fig.S2a in \cite{SM}).
In the biological context $v\sim1\mu m/sec$ for molecular 
motor assisted transport and $R\sim 5\mu m$ for T-cells, so that the
parameters in Fig. 2b correspond to $D=0.016\mu m^2/sec$ and $d=0.26\mu m$,
which is typical for vesicles. The optimal homogeneous 
strategy according to Fig. 2b is $\tilde{k}_{\rm opt}\approx 40$
and $\tilde{k}_{\rm opt}'\approx 25$, corresponding to
$k_{\rm opt}\approx8sec^{-1}$ and $k_{\rm opt}'\approx5sec^{-1}$,
which is compatible with experimental observations \cite{Loverdo-etal,Alberts}.

Again we take the optimal values 
$\tilde k_\text{opt}(\tilde D)$, $\tilde k'_\text{opt}(\tilde D)$
from $\tilde{\Delta}= 1$ and calculate with these rates the MFPT 
for the inhomogeneous directional distribution for fixed values of outward transport 
probability $p$ as a function of the cortex width $\tilde{\Delta}$. The result,
shown in Fig.\ref{Fig3}c, demonstrates that again a thin cortex 
$\tilde{\Delta}\ll 1$ yields
a much smaller search time for $p=1/2$. 
Although the target is located with a high
probability close to the center ($w= 0.75$ and $0.9$) 
the MFPT is not minimized for small values of $p$, 
corresponding to preferential radial transport towards 
the center, but for $p$ close to $1/2$. 
Similarly for fixed small cortex width $\tilde{\Delta}= 0.1$
the MFPT is minimized for $p$ close to $1/2$ 
even for large $w$ and also for fixed non-optimal 
rates $\tilde{k}$, $\tilde{k'}$ as shown
in Fig.\ref{Fig3}d.
These results are confirmed for larger values of the parameters: 
$\tilde D=0.01$ and $\tilde d=0.1$ (see Fig.S2b and Fig.S2c in \cite{SM}).

{\bf Reaction-escape problem:} 
Finally we consider the combination of the reaction and escape problem, 
where cargo has first to bind to a reaction partner before it 
can be delivered or dock at a specific area on the cell boundary
as, for instance, a synapse. A prominent example is the 
docking of lytic granules at the immunological synapse of 
cytotoxic T-lymphocytes that requires the pairing with CD3 
endosome beforehand \cite{Pairing}. 

\begin{figure}[t]
\includegraphics[width=\columnwidth]{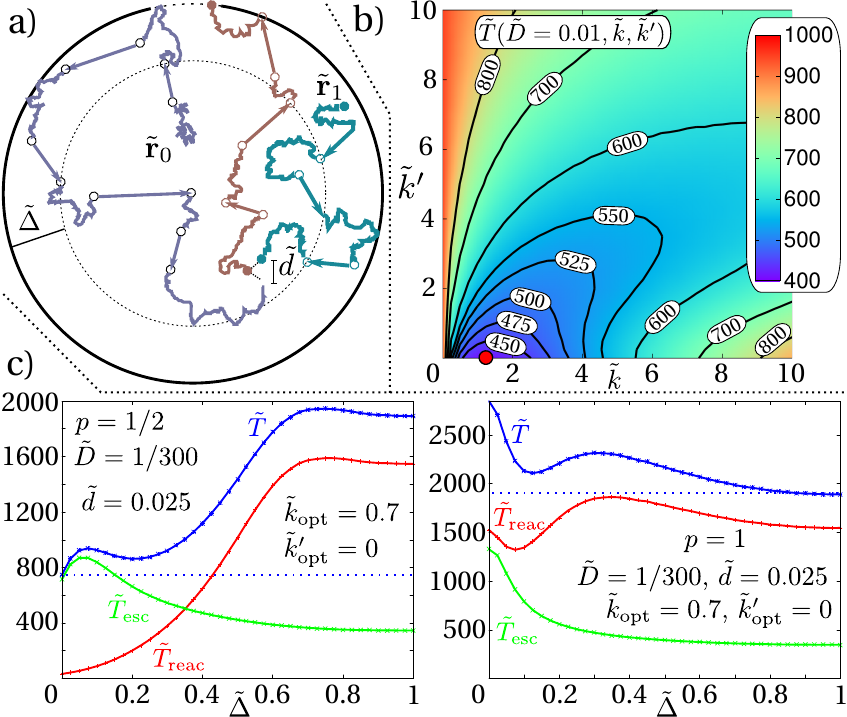}
\caption{
{\bf{a)}} 
Sketch of the reaction-escape process, involving 
a searcher and a target, both starting diffusively at random positions ${\bf \tilde{r}_0}$ (grey 
trajectory) and ${\bf \tilde{r}_1}$ (green trajectory). 
Absorption at the narrow escape region is only possible for 
searcher-target pair, and
the two particles react when coming closer than a distance $\tilde d$ 
and form a pair (brown trajectory), which will be absorbed at the 
escape region represented by the dotted segment on the cell boundary.
{\bf{b)}} Homogeneous directional distribution ($\tilde\Delta=1$): 
$\tilde T$ as a function of $\tilde k$ and $\tilde k'$ for $\tilde D=0.01, \tilde d=0.1, \vartheta_\text{abso}=\text{arcsin}(1/7)$. 
The red dot represents the minimum. 
{\bf{c)}} Inhomogeneous directional distribution:
$\tilde T_\text{reac}$, $\tilde T_\text{esc}$, $\tilde T$ for the
optimal rates $\tilde k_\text{opt}$, $\tilde k' _\text{opt}$ as a function of $\tilde{\Delta}$
for symmetric radial microtubule transport  ($p=1/2$) in the left panel,
and purely outward transport microtubule transport  ($p= 1$) in the 
right panel for $(\tilde D=1/300, \tilde d=0.025)$.
} 
\label{Fig4}
\end{figure}

Fig.\ref{Fig4}a shows a sketch of the process,
now involving two particles, a searcher and a target, each one
executing a random intermittent motion described by a
Fokker-Planck equation (\ref{FP-eq}) with additional boundary conditions:
The searcher and the target react and build a pair once they get closer than
a distance $\tilde d$ and both being in the diffusive mode.
$\tilde T_\text{reac}$, $\tilde T_\text{esc}$, and
$\tilde T= \tilde T_\text{reac} + \tilde T_\text{esc}$ denote the 
MFPT for the reaction event, the MFPT for the pair escape, and the
total MFPT, respectively.

Fig.\ref{Fig4}b shows $\tilde T$ for the homogeneous directional distribution 
($\tilde{\Delta}= 1$) as a function of the rates $\tilde k, \tilde k'$ for 
$\tilde D= 0.01, \tilde d= 0.1, \vartheta_\text{abso}= \text{arcsin}(1/7)$.
($\tilde T_\text{reac}$ and $\tilde T_\text{esc}$ are shown separately in 
Fig.S3a in \cite{SM}).
$\tilde{T}_{\rm reac}$ is minimized 
by a non-vanishing rate $\tilde{k'}$, whereas 
$\tilde{T}_{\rm esc}$ is again minimized for $\tilde{k'}=0$. 
$\tilde{T}$ is also minimized by $\tilde{k'}= 0$.
It should be noted that for a purely diffusive target the total MFPT 
is minimized by a non-vanishing rate $\tilde{k'}>0$.
In Fig.\ref{Fig4}c we show the MFPTs for the inhomogeneous directional distribution eq. (\ref{inhom-density}) as a function of the cortex width $\tilde{\Delta}$,
with the optimal rates for the homogeneous case for  $\tilde{D}= 1/300$ 
for $p= 1/2$ (outward and inward radial transport equally probable)
and for $p= 1$ (only outward radial transport). 
The data show that $p= 1/2$
and a small cortex width $\tilde{\Delta}$ 
reduce the total MFPT by 70\% in comparison with the homogeneous 
strategy $\tilde{\Delta}=1$.
For $\tilde{D}> 0.01$
one can reduce $\tilde{\Delta}$ down to zero without 
changing $\tilde T$  substantially (see Fig.S3b in \cite{SM}).
Even if one fixes the
attachment and detachment rates to non-optimal values, 
$\tilde T$ is again reduced by at least 50\% for
small cortex widths (see Fig.S3c in \cite{SM}).

To conclude we have studied for the first time the efficiency 
of spatially inhomogeneous intermittent search strategies defined by
the directional distribution $\rho_\Omega({\bf r})$ in the Fokker-Planck equation (\ref{FP-eq}) together with the rates $k$, $k'$.
For the narrow escape problem 
it turns out to be a superior strategy to allow only
radial outward ballistic transport from the center 
towards a thin sheet of thickness $\Delta$ underneath the boundary, 
where ballistic transport in all directions
is possible. This thin boundary layer allows an accelerated 
random motion along the boundary to find the escape region,
somewhat reminiscent of purely 
diffusive search with an accelerated 
surface mediated diffusion \cite{PRL-surfacemediated-Diff}.
A similar result holds for the reaction kinetics problem, in which 
the target is not located on the boundary: here again optimal 
strategies with small thickness $\Delta$ exist, in particular
better than the homogeneous strategy \cite{Loverdo-etal}, but the optimal
probability for forward/backward radial transport is now around $p=1/2$.
This result is reminiscent of an 
acceleration of purely diffusive search kinetics
by following boundaries 
with an increased diffusivity \cite{PRL-followingboundaries-Calandre}.
The reaction-escape problem combines both scenarios and 
the optimal forward/backward radial transport probability
depends on the size ratio of target and escape region.
In passing we note that qualitatively similar results 
are obtained when the searcher can find the target 
also during the ballistic motion \cite{hafner}. 

The parameters for the optimal strategies that we find 
are biologically relevant, as discussed in \cite{Loverdo-etal}
and \cite{SM}. For intra-cellular reaction kinetics 
our results therefore suggest that
the spatial organization of the cytoskeleton of cells with
a centrosome does not only minimize the characteristic time 
necessary for various random search tasks but does so 
in an economic way: instead of supporting a resource demanding
isotropic homogeneous filament network it is sufficient, and 
almost always even more efficient, to establish just
a thin actin cortex underneath the cell membrane.

\begin{acknowledgements}
This work was financially supported by the German Research Foundation 
(DFG) within the Collaborative Research Center SFB 1027. 
\end{acknowledgements}

\bibliographystyle{unsrt}

\begin{thebibliography}{10}
\bibitem{remark3}
Results do not change significantly if the searcher starts in the
ballistic mode since the time that the searcher stays in the diffusive 
mode (maximally until reaching the boundary where an automatic 
switch to the diffusive occurs) is much smaller than the overall MFPT.
\bibitem{remark}
If both reaction partners are mobile an obvious suitable strategy 
would be to transport both of them towards the MTOC to bring them 
into contact.
\end{thebibliography}

\vfill
\eject

\begin{widetext}

{\bf\large Supplementary material}
\setcounter{equation}{0}
\setcounter{figure}{0}
\setcounter{table}{0}
\makeatletter
\renewcommand{\theequation}{S\arabic{equation}}
\renewcommand{\thefigure}{S\arabic{figure}}
\renewcommand{\bibnumfmt}[1]{[S#1]}
\renewcommand{\citenumfont}[1]{S#1}

\begin{center}
\begin{figure}[!h]%
\includegraphics[width=8cm]{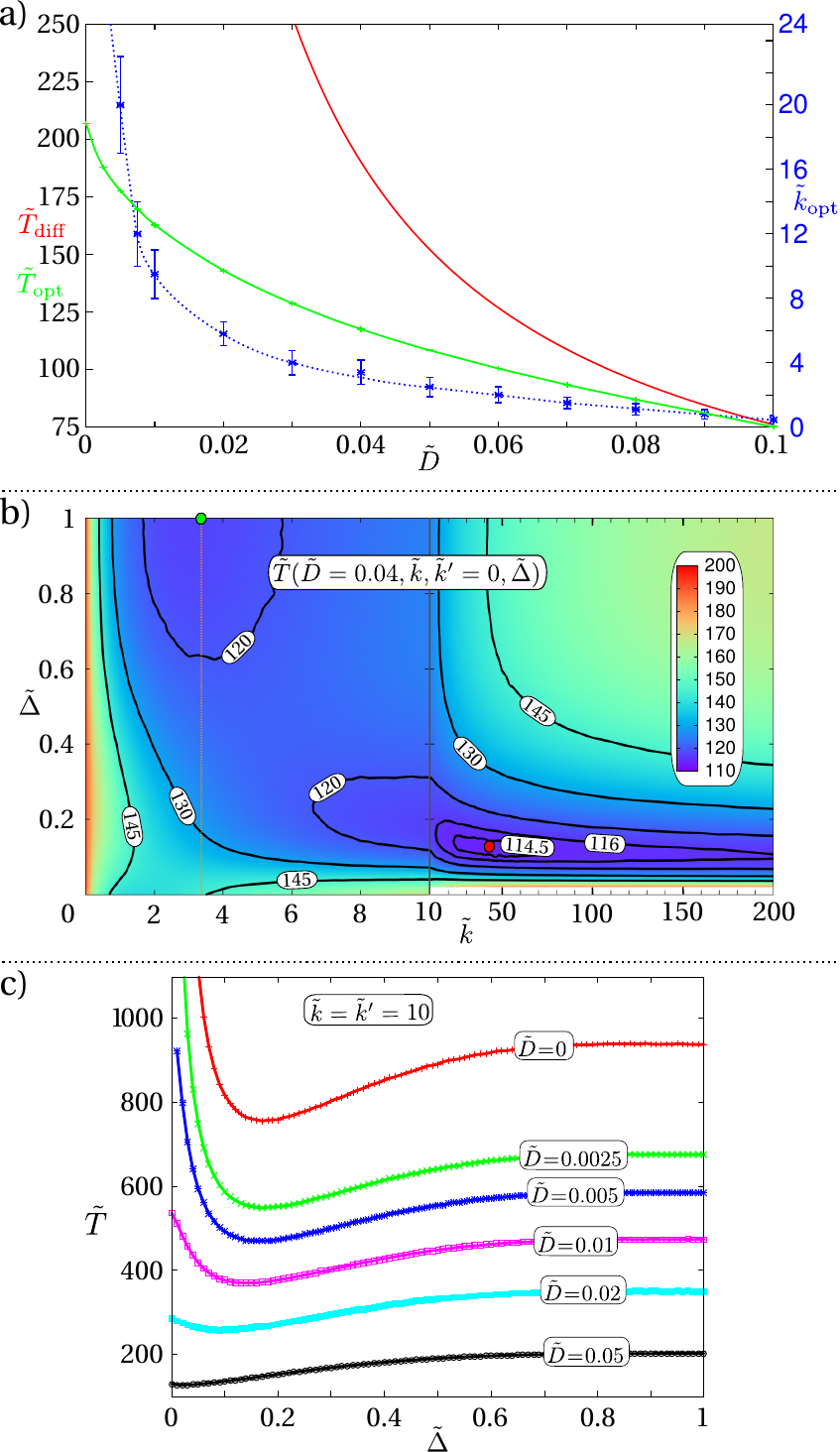}
\caption{
{\textbf{Narrow escape problem.}} {\bf{a)}} 
MFPT for the purely diffusive, ${\tilde T}_{\rm diff}$ (red line),
and intermittent search, ${\tilde T}_{\rm opt}$ (green line), 
with a homogeneous filament density 
($\tilde{\Delta}=  1$) with optimal transition rates, as functions of the
diffusion constant $\tilde{D}$. The optimal attachment rate 
$\tilde k_\text{opt}(\tilde D)$ is represented by the blue line
(and right y-axis), the optimal detachment rate is ${\tilde{k}'}_\text{opt}=  0$. 
{\bf{b)}} MFPT as a function of $\tilde k$ and $\tilde \Delta$ for $\tilde D=  0.04$, $\tilde{k}'=  0$ and $p=  1$.
The red dot indicates the global minimum, and
the green dot (top left) indicates the minimum for the homogeneous case ($\tilde\Delta=  1$). 
{\bf{c)}} $\tilde T$ as a function of $\tilde \Delta$ for fixed rates 
$\tilde k=  \tilde k'=  10$ for different $\tilde D$ and $p=  1$. 
}
\label{FigS1}
\end{figure}

\vfill\eject

\begin{figure}[!ht]
\includegraphics[width=10cm]{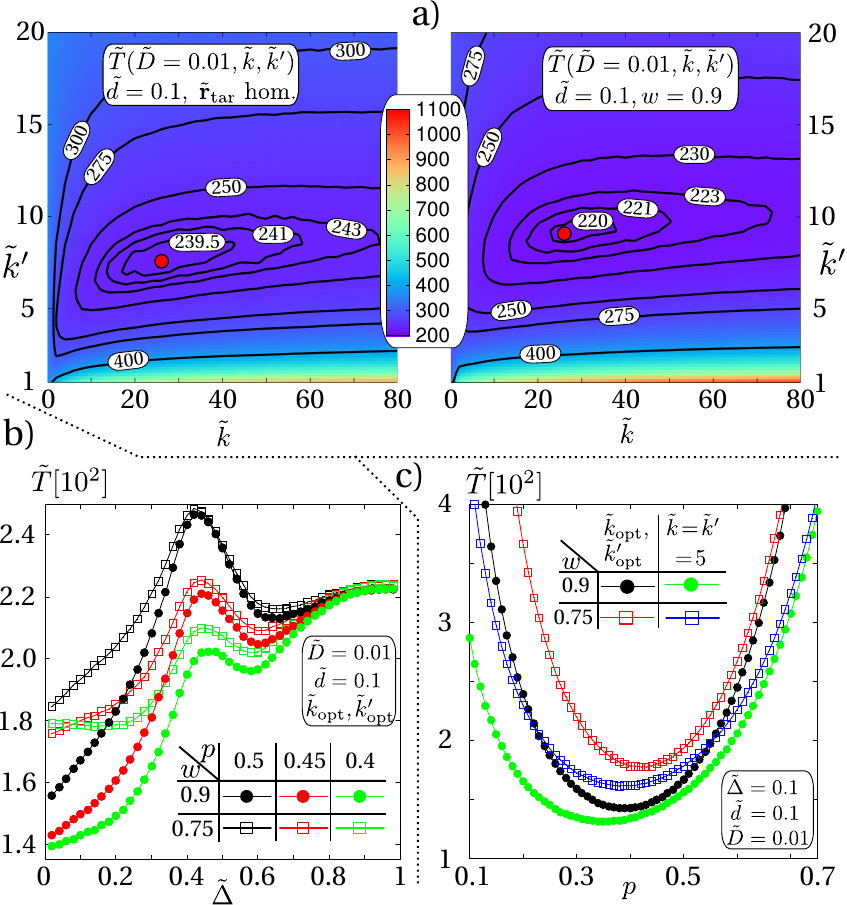}
\caption{
{\textbf{Enhanced reaction kinetic.}} 
{\bf{b)}} Homogeneous filament density:
MFPTs as a function of $\tilde k$ and $\tilde k'$ for a spatially homogeneous filament density ($\tilde{\Delta}=1$), for
({\it left}) $\tilde D=0.01, \tilde d=0.1$, homogeneously distributed target position $\tilde {\bf r}_\text{tar}$; 
({\it right}) $\tilde D=0.01, \tilde d=0.1$, 
 target position $\tilde {\bf r}_\text{tar}\leq 0.5$ with probability $w=0.9$
{\bf{c)}} MFPT for the inhomogeneous filament density 
with the optimal rates $\tilde k_\text{opt}(\tilde D, \tilde d)$
from the homogeneous case $\tilde{\Delta}=1$ as function of 
$\tilde{\Delta}$ for different values of the forward radial transport $p$
and different target positions close to the center $w$.
{\bf{d)}} MFPT as in c) but now with fixed width $\tilde{\Delta}=0.1$ 
as function of the forward probability $p$ for different 
fixed rates $k$ and $k'$ different values of $w$.
}
\label{FigS2}
\end{figure}

\vfill\eject

\begin{figure}[!h]
\includegraphics[width=10cm]{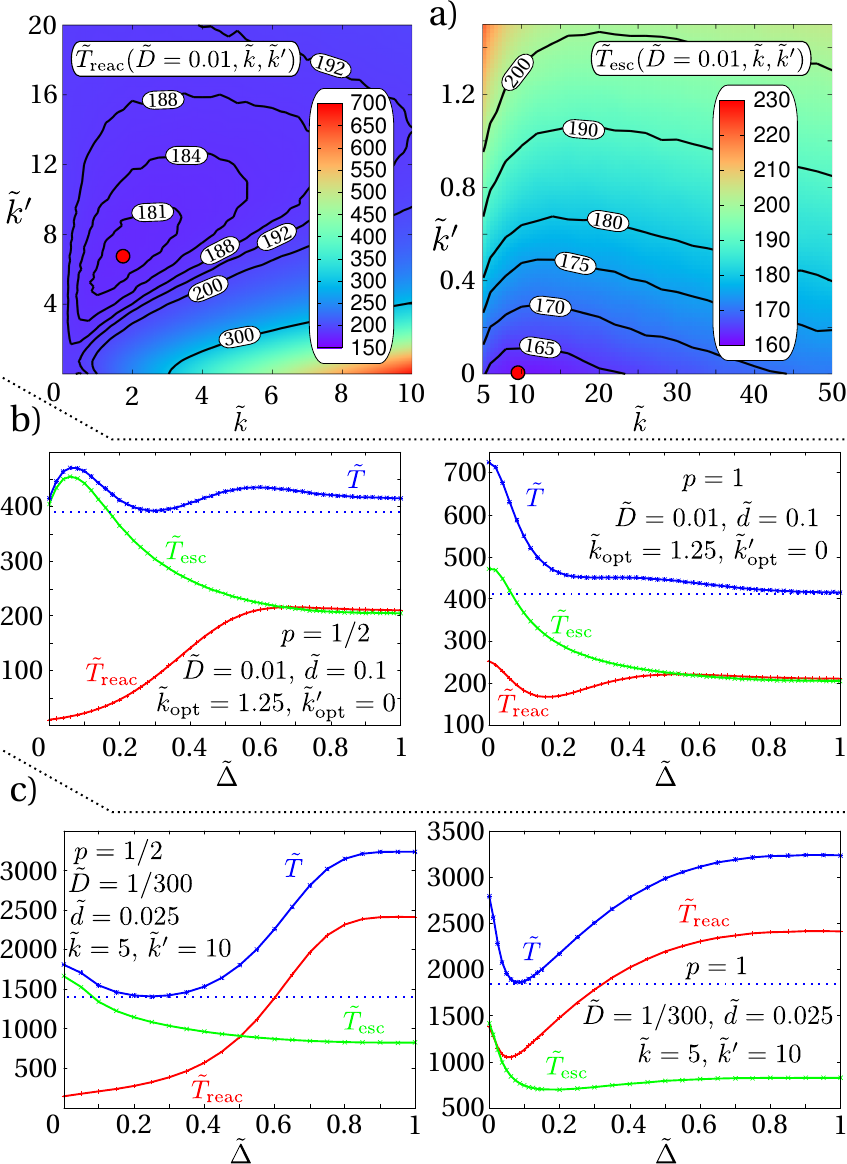}
\caption{
{\bf{a)}} Homogeneous filament density ($\tilde\Delta = 1$): 
MFPTs for reaction ($\tilde T_\text{reac}$, left), pair escape 
($\tilde T_\text{esc}$, right) as a function of $\tilde k$ and $\tilde k'$ for $\tilde D=0.01, \tilde d=0.1, \vartheta_\text{abso}=\text{arcsin}(1/7)$. 
The red dots represents the minimum. 
{\bf{b)}} Inhomogeneous filament density:
The MFPTs $\tilde T_\text{reac}$, $\tilde T_\text{esc}$, $\tilde T$ for the
optimal attachment / detachment rates ($k$, $k'$) from figure Fig.2b)
as a function of the actin cortex width ($\tilde{\Delta}$)
for symmetric radial microtubule transport  ($p=1/2$) in the left panel,
and purely outward transport microtubule transport  ($p=1$) in the 
right panel.
{\bf{c)}} The MFPTs $\tilde T_\text{reac}$, $\tilde T_\text{esc}$, $\tilde T$ for fixed, non-optimal 
attachment / detachment rates $\tilde k=5, \tilde k'=10$
for $(\tilde D=1/300, \tilde d=0.025)$.
}
\label{FigS3}
\end{figure}
\end{center}

\vfill\eject

{\large\bf Biologically relevant parameters}
\baselineskip=18pt

Optimal homogeneous search strategies for the enhanced reaction kinetics in 
biological cells have been studied in [6], where also the biological 
relevance of the parameters and the resulting optimal homogeneous 
strategies have been discussed. For heavy cargo, like vesicles 
(diameter ca. 100 $nm$) and other organelles, the diffusion constant 
is around or even below $10^{-2} \mu m^2/sec$ and therefore molecular motor assisted transport and intermittent search strategies are essential.
Consequently in this type of cargo one also finds optimal search 
strategies [6].
For light cargo like small to intermediate proteins intermittent 
search strategies are not efficient in intracellular reaction kinetics 
– and this is also what we obtain for all other search problems considered: 
for large diffusion constants purely diffusive search is always optimal 
(see Fig S1). 

In [6] it was argued that for plausible values of 
the velocity ($v\sim 1\mu m/sec$) and diffusion constants for
vesicles ($\sim 10^{-2} \mu m^2/sec$) the optimal attachment 
and detachment rates turned out to be of the order of 
$k,k'\sim 10 sec^{-1}$ for reaction radii $d\sim 0.1\mu m$ 
and thus compatible with experimental observations [25].

For the inhomogeneous search strategies we take the optimal 
rates for the homogeneous case, which were already found to be 
biologically relevant [6], and vary the cortex width $\Delta$ 
and the radial forward probability $p$. The width of the 
actin cortex of cells is of the order of 0.2-0.4 $\mu m$,
and the radial forward probability depends on the equipment
of cargo with the forward and backward motors, kinesin and dynein, 
respectively. The optimal widths we find vary from case to case
between zero and 0.2 $R$, where $R$ is the cell radius. 
When considering T-cells the radius is around 5 $\mu m$.
In the following we fix $R=5\mu m$ and $v=1\mu m/sec$ and
consider the three search problems treated in the main 
text separately.

{\bf Narrow escape problem:}
As shown in Fig. S1a the optimal attachment rate 
$\tilde{k}_{\rm opt}$ 
for the homogeneous search strategy decreases
with increasing diffusion constant $\tilde{D}$. 
With $R=5\mu m$, $v=1 \mu m/sec$ and $D=0.05 \mu m^2/sec$ 
one has $\tilde{D}=0.02$ and from Fig. S1a 
$\tilde{k}_{\rm opt}\approx 10$ and thus $k_{\rm opt}\approx 2 sec^{-1}$.
For smaller diffusion constants $k_{\rm opt}$ increases further.
In the inhomogeneous case we take these optimal values, as shown in 
Fig. 1c in the main text and show the MFPT as a function of the
cortex width $\Delta$. For diffusion constants smaller than $\tilde{D}=0.02$,
i.e.\ $D<0.1\mu m^2/sec$ there is a minimum, which moves towards $\tilde{\Delta}\approx 0.1$ for further decreasing diffusion constant.
$\tilde{\Delta}\approx 0.1$ corresponds to $\Delta\approx0.5\mu m$,
close to the actual width of the actin cortex in cells.
Fig. S1c shows that one can also fix the detachment and attachment 
rates to biologically relevant values (here $k=k'=2 sec^{-1}$)
and one still gets a minimal MFPT for small cortex widths.

{\bf Enhanced reaction kinetics:}
The optimal switch rates for the homogeneous search for
$\tilde{D}=1/300$ and reaction distance $\tilde{d}=0.025$,
corresponding to $D=0.016 \mu m^2/sec$ and $d=0.125 \mu m$,
are according to Fig. 2b $\tilde{k}_{\rm opt}\approx 40$
and $\tilde{k}_{\rm opt}'\approx 25$, which corresponds to
$k_{\rm opt}\approx 8 sec^{-1}$ and $k_{\rm opt}'\approx 5 sec^{-1}$.
For the inhomogeneous case Fig. 2c shows that for $p=0.5$
(equal probability for forward and backward radial transport) 
a thin cortex yields the smaller MFPTs (actually the minimum is
at $\Delta=0$). Fig. S2 yields similar numbers.

{\bf Reaction-escape problem:}
Fig. S3c shows again for the inhomogeneous case that for switch rates 
fixed to biologically relevant values (here $k=0.5k'=1 sec^{-1}$) 
a thin cortex yields smaller MFPTs than the homogeneous search 
strategies (minimum smaller than $\tilde{\Delta}=0.1$, 
i.e. $\Delta<0.5\mu m$).

\end{widetext}

\end{document}